\documentclass[a4paper,usenatbib,fleqn]{mnras}

\makeatletter

\newcommand{\Rmnum}[1]{\expandafter\@slowromancap\romannumeral #1@}
\makeatother

\usepackage{graphicx}   
\usepackage{amssymb}    
\usepackage{subcaption}
\usepackage{color,soul}
\usepackage{xcolor}
\usepackage{newtxtext,newtxmath}
\usepackage{float}
\restylefloat{table}
\usepackage{dblfloatfix}
\usepackage{booktabs}
\usepackage{multirow}
\usepackage{array}
\usepackage{amsmath}    

\usepackage{color}

\def\prd{{\rm Phys. Rev. D}}
\def\apj{{\rm ApJ}}

\def\mnras{{\rm MNRAS}}

\def\hmpc{h^{-1}{\rm Mpc}}
\def\hgpc{\;h^{-1}{\rm Gpc}}

\def\lya{Ly$\alpha$ }

\def\simlt{\lower.5ex\hbox{$\; \buildrel < \over \sim \;$}}
\def\simgt{\lower.5ex\hbox{$\; \buildrel > \over \sim \;$}}

\title[]{
 Weak Lensing the non-Linear \lya Forest
}

\author[Shaw et al.]{\parbox{18cm}{Patrick Shaw$^{1,2}$\thanks{E-mail: pshaw2@andrew.cmu.edu}, 
Rupert A.C. Croft$^{1,2}$,
and R. Benton Metcalf$^{3,4}$
}\vspace{0.3cm}
\\
$^{1}$ McWilliams Center for Cosmology, Dept. of Physics, 
Carnegie   Mellon  University, Pittsburgh, PA 15213, USA
\\
$^{2}$ NSF AI Planning Institute for Physics of the Future, 
Carnegie   Mellon  University, Pittsburgh, PA 15213, USA
\\
 $^{3}$ Dipartimento di Fisica \& Astronomia, Universit\'{a} di Bologna, via Gobetti 93/2, 40129 Bologna, Italy
\\
$^{4}$ INAF-Osservatorio Astronomico di Bologna, via Ranzani 1, 40127 Bologna, Italy
\\
}

\begin{document}

\topmargin=-1.0cm
\maketitle


\begin{abstract}
We evaluate the performance of the Lyman-$\alpha$ forest weak gravitational lensing estimator of Metcalf et al. on forest data from hydrodynamic simulations and ray-trace simulated lensing potentials. 
We compare the results to those obtained from the Gaussian random field simulated \lya forest data and lensing potentials used in previous work. We find that the estimator is able to reconstruct the lensing potentials from the more realistic data, and investigate dependence on spectrum signal to noise. The non-linearity and non-Gaussianity in this forest data arising from gravitational instability and hydrodynamics causes a reduction in signal to noise by a factor of $\sim2.7$ for noise free data and a factor of $\sim 1.5$ for spectra with signal to noise of order unity (comparable to current observational data). Compared to Gaussian field lensing potentials, using ray-traced potentials from
N-body simulations incurs a further signal to noise reduction of a factor of $\sim1.3$ at all noise levels. The non-linearity in the forest data is also observed to increase bias in the reconstructed potentials by $5-25\%$, and the ray-traced lensing potential further increases the bias by $20-30\%$. We demonstrate methods for mitigating these issues including Gaussianization and bias correction which could be used in real observations.

\end{abstract}


\section{Introduction}
\label{intro}
Weak gravitational lensing is the process by which the gravitational field sourced by foreground matter minimally deflects the observed images of background light sources. In contrast to strong gravitational lensing or microlensing, the optical distortions are small and can only be detected through statistical methods. The weak lensing signatures contained in observations of various continuous fields can be used to gain novel information about the foreground matter distribution, making weak lensing a valuable cosmological probe (\citealt{lensingreview}). In the case of a continuous field, weak lensing causes distortions in the expected statistics of the observed field. These distortions can be quantified and used to reconstruct properties of the foreground matter (e.g., \citealt{cmbestimator}). To this end, estimators have been developed for various continuous fields including the Cosmic Microwave Background (CMB) (\citealt{bernard97,1997ApJ...489....1M,1998ApJ...492L...1M,zald99,huok02,schaan19}) and 21 cm line radiation (\citealt{madau97,furl06}). In this work we focus on extending the techniques of weak lensing to a novel source field, the Lyman-alpha forest (\lya forest).

The \lya forest is a set of absorption features observed in the spectra of high redshift ($z\sim3$) galaxies and quasars (see reviews by e.g., \citealt{rauch98,prochaska19}). Light from these background sources is redshifted and absorbed by the intervening neutral hydrogen density field at the wavelength corresponding to the \lya transition. The result is that a one-dimensional sampling of the hydrogen density field along the line of sight can be obtained from the spectrum of a background source. If many background sources are observed, the sampled ``skewers'' can be combined to produce a three-dimensional map of the hydrogen density field. The \lya forest is an ideal candidate for weak lensing as it is well understood and easily simulated, has large amounts of observational data available or soon to be available (\citealt{LATIS,CLAMATO,eBOSS}), and is sensitive to lower redshifts than CMB lensing ($z\sim1$ compared to $z\sim2.5$ for the CMB, \citealt{bennoise,CMBkernel}). However, the \lya field is sparsely sampled as the positions of observable background sources are irregular. This poses a challenge as the Fourier space-based techniques employed in CMB and 21 cm weak lensing fail for a field that is not regularly and fully sampled. To this end, \cite{ben1} have derived an estimator for the foreground lensing potential that is suitable for the sparse geometries of \lya observations.

In this paper, our goal is to further develop \lya forest weak lensing techniques by testing the \lya estimator derived in \cite{ben1} with more realistic \lya forest and lensing simulations. The tests in \cite{ben1} use simulations that assume that both the \lya forest and the foreground lensing potential are Gaussian random fields (GRF). This assumption should hold at larger scales, but not at smaller scales (e.g., \citealt{bbks} 
). It is important to understand what impact the non-linearity introduced by more realistic hydrodynamic forest simulations has on the performance of the estimator as it will be present in real observational data. To this end, the paper will be organized as follows: first we will briefly introduce the weak lensing formalism and the estimator used, then we will describe our methods for simulating data, and finally we will evaluate the impacts more realistic data have on the estimator's performance.  


\section{Reconstruction Method}
\label{method}
\subsection{Lensing formalism}
In this work we reconstruct the foreground lensing potential from the statistical distortions observed in simulated \lya flux data. In the Born and thin lens approximation, an observed pixel image will be deflected on the sky by an angle $\vec{\alpha}( \vec{\theta})$  according to
\begin{align}
\label{deflection}
    \vec{\alpha}(\vec{\theta})=\frac{1}{\pi} \int_{\mathbb{R}^{2}} \mathrm{~d}^{2} \theta^{\prime} \kappa\left(\vec{\theta^{\prime}}\right) \frac{\vec{\theta}-\vec{\theta}^{\prime}}{\left|\vec{\theta}-\vec{\theta}^{\prime}\right|^{2}}.
\end{align}
$\kappa(\vec{\theta})$ is the dimensionless convergence defined as 
\begin{align}
    \kappa(\vec{\theta})=\frac{3}{2} \frac{\Omega_{m} H_{0}^2}{c^{2}} \int_{o}^{\chi_{s}} d \chi\left[\frac{d_{A}(\chi) d_{A}\left(\chi, \chi_{s}\right)}{d_{A}\left(\chi_{s}\right)a}\right] \delta(\boldsymbol{\theta}, \chi),
\end{align}
where $\delta(\vec{\theta}, \chi)$ is the density contrast at a given radial coordinate, $\chi$ is the comoving distance, $d_{A}(\chi)$ is comoving angular size distance, $H_0$ is the Hubble constant, $\Omega_m$ is the matter density parameter, and $a$ is the scale factor. The gradient of the lensing potential yields the deflection field
\begin{align}
    &\vec{\alpha}(\vec{\theta})=\nabla\phi(\vec{\theta}),
\end{align}
and is related to the convergence by a Poisson equation
\begin{align}
    &\nabla^{2} \phi(\vec{\theta})=2 \kappa(\vec{\theta}).
\end{align}
Therefore, the potential can be obtained from the convergence according to 
\begin{align}
    \label{potfromkappa}
    \phi(\vec{\theta})=\frac{1}{\pi} \int_{\mathbb{R}^{2}} \mathrm{~d}^{2} \theta^{\prime} \kappa\left(\vec{\theta^{\prime}}\right) \ln \left|\vec{\theta}-\vec{\theta^{\prime}}\right|.
\end{align}
\subsection{Quadratic estimator}
We reconstruct the lensing potential using the quadratic estimator derived by \cite{ben1}. The estimator reconstructs the amplitudes of the Legendre polynomial expanded potential,
\begin{align}
    \phi(\boldsymbol{\theta})=\sum_{m=0}^{N_{x}} \sum_{n=0}^{N_{y}} \hat{\phi}_{mn} P_{m}(x) P_{n}(y).
\end{align}
The $P_n$ are the Legendre polynomials, and the variables are scaled such that
\begin{align}
x \equiv 2 \left(\frac{\theta_{1}-\theta_{1}^{o}}{\Delta \theta_{x}}\right)-1, \quad y \equiv 2 \left(\frac{\theta_{2}-\theta_{2}^{o}}{\Delta \theta_{y}}\right)-1
\end{align}
where the $\left(\theta_1,\theta_2\right)$ are the angular coordinates of the field origin (lower left) and the $\Delta \theta_{x,y}$ are the field widths. The estimate for the parameters $\hat{\phi}_\mu$ is given by
\begin{align}
    \hat{\phi}_{\mu}=\frac{1}{2} F_{\mu \nu}^{-1}\left(\boldsymbol{\delta}^{\top} \mathbf{C}^{-1} \mathbf{P}^{* \nu} \mathbf{C}^{-1} \boldsymbol{\delta}-\rm{tr}\left[\mathbf{C}^{-1} \mathbf{P}^{* \nu}\right]\right),   
\end{align}
where $F_{\mu\nu}^{-1}$ is the inverted Fisher matrix, $\boldsymbol{\delta}$ is a vector of the \lya flux overdensities, $\mathbf{C}$ is the covariance matrix between \lya flux pixels including intrinsic correlations and noise, and $\mathbf{P}$ is constructed from the derivatives of the chosen basis functions. This discretized estimator works for the sparse geometry of the \lya forest, in contrast to the Fourier-based methods employed in continuous field lensing such as the CMB (\citealt{cmbestimator}).

It is important to note that this estimator requires a priori knowledge of the \lya flux field correlation function. Errors in the assumed correlation function will lead to bias in the estimator. In this work we use the model proposed in \cite{McDonald} to estimate the \lya power spectrum from which we compute the \lya pixel correlation function (see Appendix A and B of \citealt{ben1} for details).
\subsection{Geometry and implementation}
\label{geometry}
In this work, we consider a $0.655 \times 0.655 \deg^2$ field with $512$ sightlines containing $512$ pixels each. This corresponds to a source density of $\eta \sim 1200 $ sources $\deg^{-2}$ which is comparable to currently available observations (LATIS currently has $\eta \sim 1600 $ sources $\deg^{-2}$ over $0.8925\deg^2$, \citealt{LATIS}, CLAMATO currently has $\eta \sim 1500$ sources $\deg^{-2}$ over $0.157\deg^2$, \cite{CLAMATO}, and DESI will have $55$ sources $\deg^{-2}$ over a much larger $14000 \deg^2$ field \citealt{DESI}).
This geometry corresponds approximately to a $50\times50\times400 ({\hmpc{}})^{3}$ volume. We focus on geometries comparable to LATIS and CLAMATO because these data are currently available and are more consistent with previous work allowing for direct comparison with other Gaussian random field tests (\citealt{ben1}). We expect smaller, higher density geometries like LATIS and CLAMATO to have more signal as there is more forest power at these scales compared to DESI. Future work will involve determining whether the larger amount of data in DESI sufficient to overcome the weaker signal. We discuss this further in Section \ref{disc} below.

The positions of the sightlines are determined by randomly populating half of the points on a $32\times32$ grid to approximate the irregular source distribution from a real \lya forest observation. While this method leads to a geometry that is less sparse than a true survey, tests with the more realistic sparse geometries described in \cite{ben1} showed a reduction in signal to noise of only $1.09$ for optimistic noise levels (pixel noise $\sigma=0.3\left<F\right>$) and $.91$ for realistic noise levels ($\sigma=0.6\left<F\right>$) in the sparse case. S/N reduction was larger for very small amounts of noise ($\sigma=0.1\left<F\right>$) and no noise with reductions of $1.8$ and $2.2$ respectively. For the realistic noise dominated cases the impact is negligible, but if surveys achieve lower noise levels this effect should be investigated. The sampling approach described here is necessary due to the geometry of the hydrodynamic \lya forest simulation sample spectra we used for our tests.  
The pixel length is $0.78 \hmpc$ compared to $\sim 1.2 \hmpc$ in LATIS.

The estimator is calculated using a C++ code. Due to the large matrix inversions involved, this computation can be expensive. In this work, the sparse \lya pixel geometry is held constant for a four pixel deep slice in redshift to mitigate this cost. This way only one estimator can be constructed and applied repeatedly to the many redshift slices comprising the total data set. The results from these bins can be combined to provide an overall estimate for the potential according to 
\begin{align}
    \hat{\phi}_{\mu} =\frac{1}{2} F_{\mu \nu}^{-1} \sum_{k=1}^{n} \tilde{\phi}_{\nu}^{k},
\end{align}
where the $\tilde{\phi}_{\nu}^{k}$ are the parameter estimates from each bin. In the case of redshift bins with constant noise and identical geometry, this expression reduces to an average over the estimates from each bin. 
 This approach is justified because the signal contribution of correlations between even shallow slices in redshift are small, as justified by  \cite{ben1}. We reconstruct up to order five in the Legendre modes in either direction, yielding $22$ reconstructed parameters. The $\left(0,0\right)$, $\left(0,1\right)$, and $\left(1,0\right)$ modes are filtered because they are not measurable from lensing. 

\section{Simulations}
Testing the efficacy of the estimator requires simulating both \lya forest data and a foreground lensing potential. In this paper we are interested in evaluating estimator performance on astrophysically realistic data sets. To this end, we will employ \lya hydrodynamic simulations and lensing potentials calculated from ray-traced N-body simulations. These will be compared to GRF simulations of the same fields.

\subsection{Gaussian \lya forest and potential}

Previous work (\citealt{rupert1,ben1,ben2}) has modeled both the lensing potential and the \lya forest as GRFs. We conduct the reconstruction process under these assumptions as a control. In the case of the forest, the correlation function for the \lya pixels is computed from the parameterised power spectrum fitting function proposed by \citealt{McDonald}. Then, the \lya pixels are simulated directly from the covariance matrix obtained using a Cholesky decomposition (see \citealt{ben1} for details). The \lya pixel correlation matrix $\mathbf{C}$ can be decomposed as
\begin{align}
   \mathbf{C}=\mathbf{L L}^{T}.
\end{align}
Therefore, our simulated \lya pixels, $\delta_i$, are given by
\begin{align}
    \delta_i = \mathbf{L}x_i,
\end{align}
where the $x_i$ are generated by sampling a standard normal distribution. The resulting pixels will have the same statistics as if they were sampled from a GRF. In this case, the correlation function assumed by the estimator and the true correlation function of the forest are equivalent by construction so we would expect the estimator to be unbiased.

Next, the foreground lensing potential is simulated. This potential is also assumed to be a GRF. Using the power spectrum calculated from CAMB (\citealt{camb}) and CosmoSIS (\citealt{cosmosis} ), a field eight times larger than the intended reconstruction field is simulated using the standard Fourier space method. This field is then cropped to the desired size, avoiding imposing periodic boundary conditions. This potential can be integrated to obtain the lensing deflection field (see Equation \ref{deflection}). In this case the \lya forest can be easily simulated for any pixel geometry, so the estimator maintains a constant pixel layout and different potentials are realized by ``undeflecting'' the flux pixel locations to what their unlensed positions would be given a particular lensing potential. The estimator then attempts to reconstruct the lensing potential using these lensed \lya pixels. The reconstructions can then be compared to the known input to evaluate performance.

\begin{figure}
    \includegraphics[width=\linewidth]{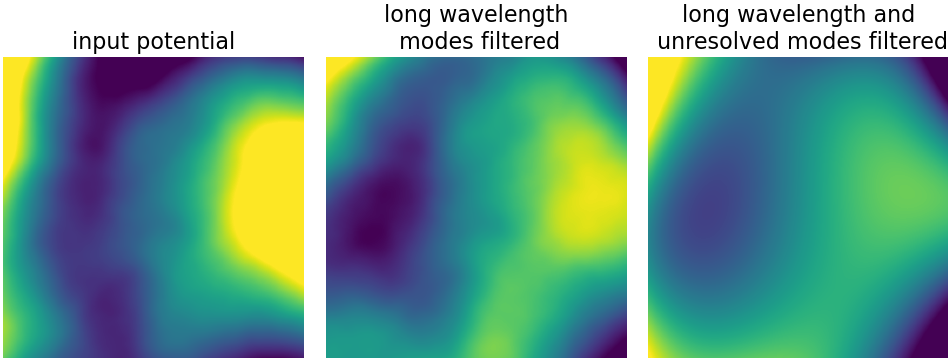}
    \caption{Demonstration of the visual effect of filtering the correlation function dependent modes and unresolved modes in a lensing potential map. The field of view is a square of side length 0.655 deg, at redshift $z=3$.}
    \label{filtered potential}
\end{figure}

\subsection{ \lya forest from hydrodynamic simulations}
\label{nglya}
We would like to compare the performance of the estimator in the case of Gaussian simulated fields to more realistic fields. First, we introduce non-linearity (and therefore non-Gaussianity) into the \lya flux field. We anticipate this will have a more marked impact than the introduction of a non-linear lensing potential. We accomplish this by using a more realistic \lya forest from a smoothed particle hydrodynamics (SPH) simulation. This simulation used the {\small P--GADGET} ( \citealt{hydrosimcode1,hydrosimcode2}) code to evolve $2 \times 4096^2 = 137$ billion particles in a $(400 \hmpc)^3$ volume at $z=3$ in $\Lambda$CDM with $h = 0.702$, $\Omega_{\Lambda} = 0.725$, $\Omega_m = 0.275$, $\Omega_b = 0.046$, $n_s = 0.968$, and $ \sigma_8 = 0.82.$ (see \citealt{hydrosim1,hydrosim2} for more details). This simulation volume yields $256\times256$ \lya sightlines which we sample with $512$ pixels each. That allows us to perform the reconstruction procedure on $64$ realizations of the field geometry described in \ref{geometry}.

In this case, the unlensed \lya fluxes are known only at fixed gridpoints. This means that the ``observed'' deflected positions will vary depending on the lensing potential used. Therefore, we construct a unique estimator from the lensed flux positions for each potential tested. In this first study the lensing potentials used remain Gaussian and are obtained in the same way as described in the previous section. 

One difficulty in the case of the non-Gaussian forest from a hydrodynamic simulation is that the correlation function is no longer known exactly. We find that two modes (the longest wavelength $\left(0,2\right)$ and $\left(2,0\right)$ modes) are particularly sensitive to the normalization of the assumed correlation function. 
Attempts were made to mitigate this by fitting the assumed correlation function to a direct estimate of the correlation function measured from the SPH \lya forest. In these calculations, the correlation function is expressed in the basis of Legendre polynomials
\begin{align}
    \xi(s, \alpha)=\sum_{\ell} \xi_{\ell}(s) P_{\ell}(\cos (\alpha)),
\end{align}
where $s$ is the absolute separation, and $s_{\|} /|\boldsymbol{s}|=\cos (\alpha)$ is the angular separation. In the present work, the amplitudes of the first two non-zero modes $\left(\xi_0,\xi_2\right)$ were fit to a direct computation of $\xi(s, \alpha)$ from the SPH simulation data. Without this fit (i.e., assuming instead the correlation function used in our linear theory simulations), we find in tests that these first two reconstructed modes can be more than an order of magnitude different from their true value.  The fit helps somewhat, but more work is required to formulate a method to match the correlation function exactly. As the other modes are reconstructed well and are not sensitive to the assumed correlation function, we filter these problematic modes in the image reconstructions and statistical measures of performance. These difficulties are separate from the issue of non-Gaussianity and should be addressed in future work. Fig.  \ref{filtered potential} shows the visual impact of filtering both these long wavelength modes and the unresolved small wavelength modes. The majority of the structure in the potential remains.   



\subsection{Ray-traced lensing potential}
We also test the introduction of more realistically simulated lensing potentials in combination with the non-linear forest. These potentials are obtained from the ray tracing simulations described in \cite{ngpot}. The matter densities used to perform the ray tracing calculations are obtained from the BigMDPL simulation (\citealt{raytracingsim}) which evolved $3840^3$ particles in a $2.5\hgpc$ box with parameters from \textit{Planck} data \cite{planck}. A convergence map is calculated by deflecting through 24 lens planes out to a source redshift of $z_\textrm{s}=2.2$. We split a $5.5\times 1.6 \deg^2$ convergence map into five fields of the desired size ($0.655 \times 0.655 \deg^2$) and then convert it into lensing potential using Equation \ref{potfromkappa}.

\subsection{Noise and varying data set size}

\label{noise}
We investigate the impact of \lya pixel noise and varying data set size to facilitate comparison with currently available observations. We work with the flux overdensity, $\delta_{\mathrm{F}}$, a quantity with zero mean.
\begin{align}
    \delta_{\mathrm{F}}=\frac{F}{\langle F\rangle}-1,
\end{align}
where $F$ is the observed \lya flux.
Noise is added in units of the mean flux $\langle F\rangle$. Three different levels of noise were considered, $0.1\langle F\rangle$, $0.3\langle F\rangle$, $0.6\langle F\rangle$. Noise is added by randomly and independently sampling a Gaussian distribution with standard deviation corresponding to the desired noise level ($0.1$, $0.3$, $0.6$) and adding the result to the simulated \lya flux pixel. For comparison, both the CLAMATO and LATIS observations have median pixel noise of $\sim0.6\langle F\rangle$. Previous work (\citealt{ben1}) focused on higher noise levels ($0.5\langle F\rangle$, $0.6\langle F\rangle$, $0.8\langle F\rangle$).  We assume that the pixel noise is Gaussian and uncorrelated. Relaxing these assumptions is left to future work. 
Most of the results we present are obtained from an average of $64$ realizations of the \lya forest, to allow more precise evaluation of biases and errors than would be possible with a single realization. However, in Fig. \ref{realobs} we also average over fewer realizations to give a qualitative sense of how effective reconstruction from a single data set could be. Because the source geometry remains unchanged between realizations, averaging over them is equivalent to lengthening the sightlines. For example, averaging over two realizations of the forest is equivalent to a single observation where each sightline is twice as long and contains twice as many pixels.

\section{Results}
\label{results}

\subsection{Evaluating estimator performance}
\label{statistics}
\begin{figure}
    \includegraphics[width=\linewidth]{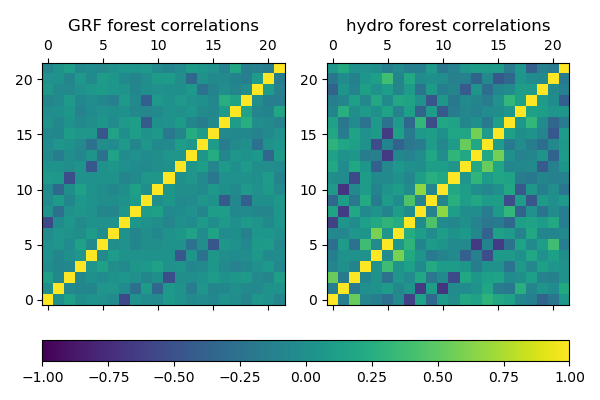}
    \caption{Normalized covariances between the 22 reconstructed mode amplitudes in the case of the Gaussian and non-Gaussian forest. Non-Gaussianity in the forest introduces correlations between modes as would be expected.}
    \label{covariances}
\end{figure}

To compare estimator performance for these different cases, we produce lensing potential reconstructions for $64$ Monte Carlo realizations of the \lya forest for five different input potentials. For each reconstruction, we perform a linear fit for the slope of the reconstructed Legendre amplitudes versus the input amplitudes. 
We then compute an error bar from the standard deviation of the distribution of slope fits for the $64$ realizations. The ratio of the average slope to the standard deviation of the slope distribution gives an estimate of the S/N for one realization of the forest. We also compute a reduced $\chi^2$ using the standard deviations of the reconstructed modes. The covariances introduced by the non-linearity  are relatively small (see  Fig. \ref{covariances}), and we found that their use added numerical instability without improving the $\chi^2$ so the amplitude standard deviations alone were used.

\begin{figure*}
    \includegraphics[width=\textwidth]{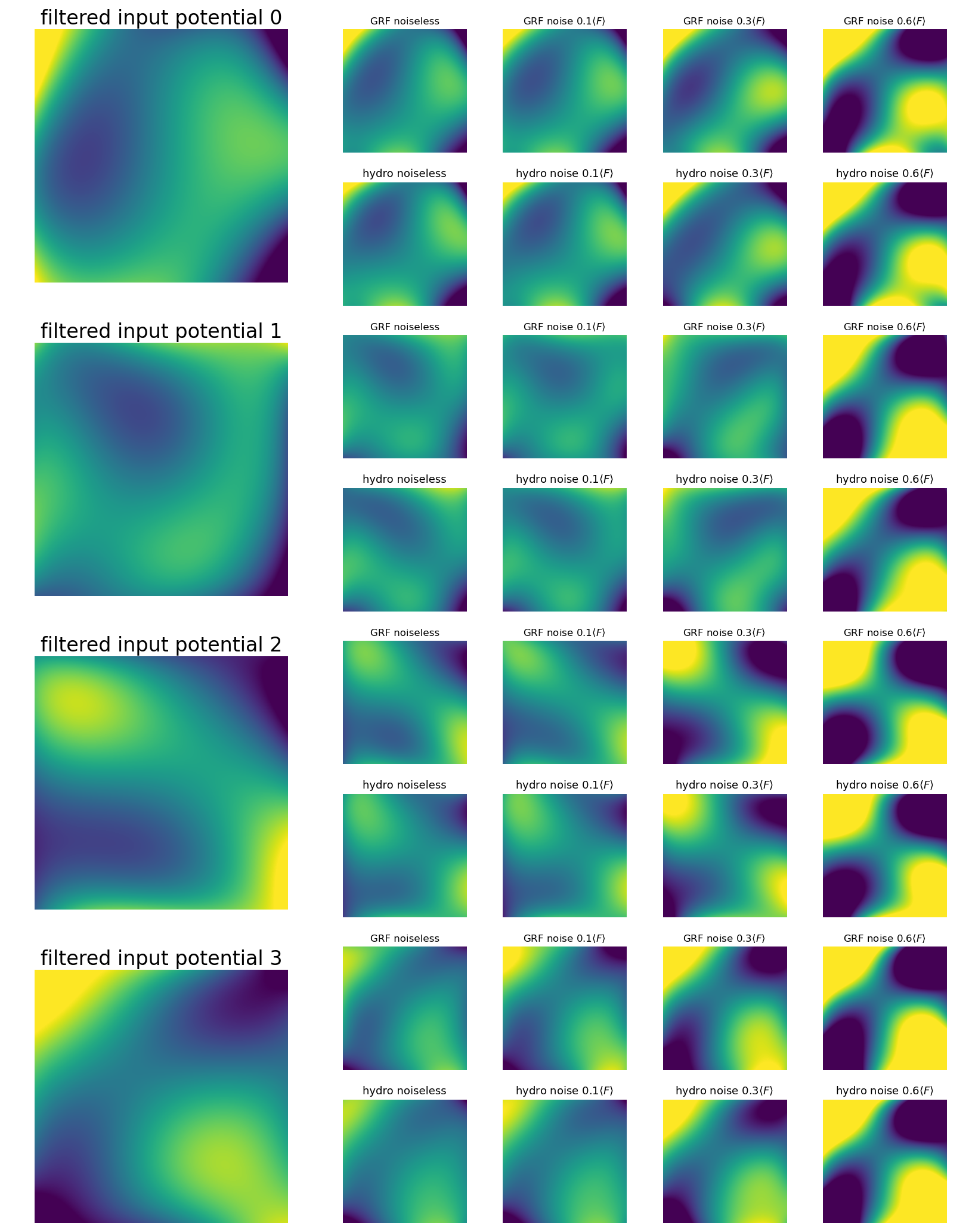}
    \caption{Lensing potential reconstructions (right panels) for four different potentials at four different noise levels for the Gaussian and non-Gaussian forest. The input potentials (left panels) have modes $\left(0,0\right)$, $\left(0,1\right)$, $\left(1,0\right)$, $\left(0,2\right)$, $\left(2,0\right)$ and modes higher than order five filtered as these modes are either unresolved or too sensitive to choice of correlation function (see Section \ref{nglya}). The reconstructions are averaged over 64 realizations of the \lya forest pixel geometry described in Section \ref{geometry} }
    \label{reconstructions}
\end{figure*}

\begin{figure*}
    \includegraphics[width=\textwidth]{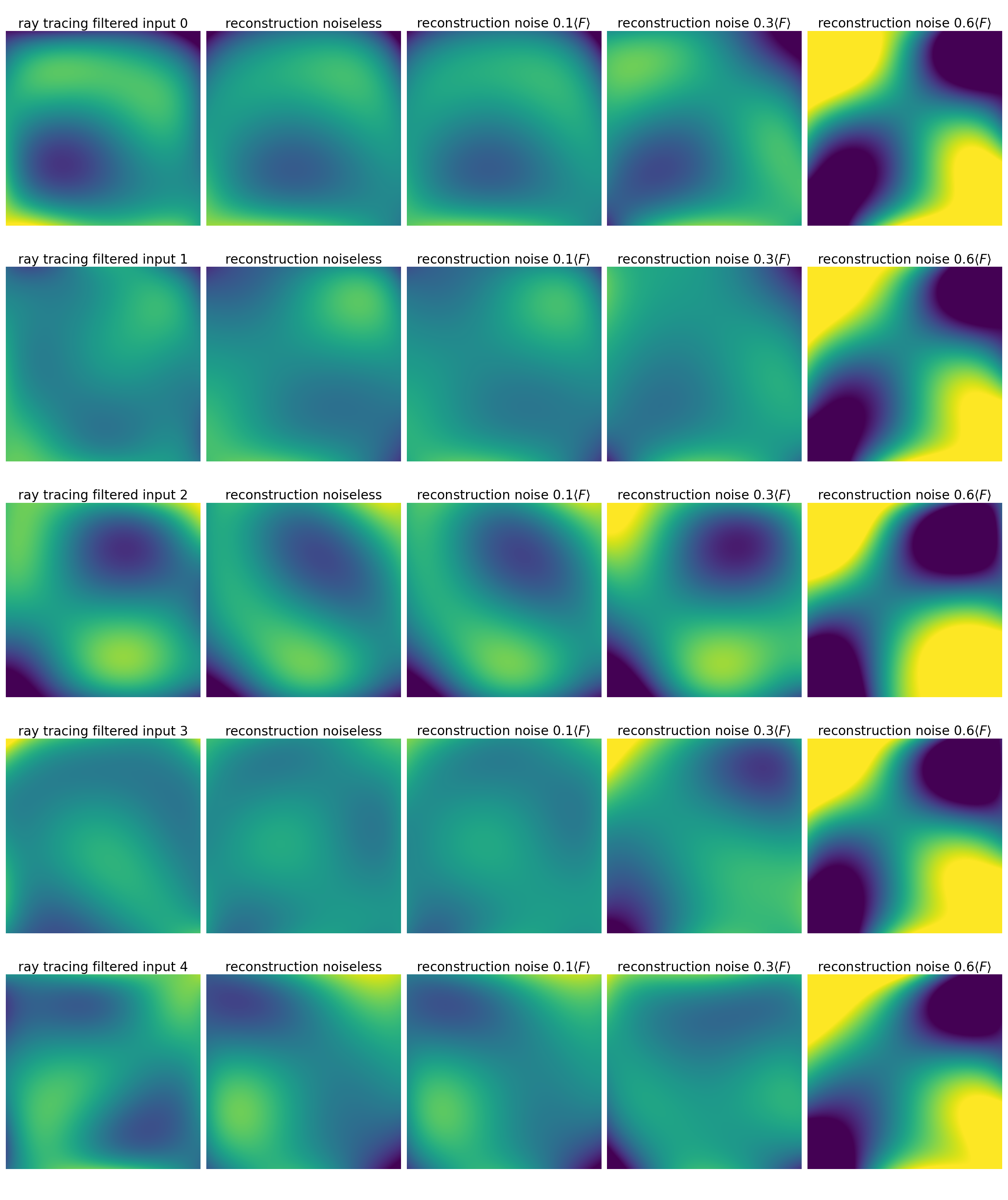}
    \caption{Lensing potential reconstructions for four different non-Gaussian potentials at four different noise levels for the non-Gaussian forest. The input potentials have the $\left(0,0\right)$, $\left(0,1\right)$, $\left(1,0\right)$, $\left(0,2\right)$, $\left(2,0\right)$ and modes higher than order five filtered as these modes are either unresolved or too sensitive to choice of correlation function. The reconstructions are averaged over 64 realizations of the \lya forest pixel geometry described in section \ref{geometry}.
    As in Figure \ref{filtered potential} the field of view is a square of side length 0.655 deg, at redshift $z = 3$.}
    \label{exampleConstQuasar}
\end{figure*}


\begin{figure*}
    \includegraphics[width=\textwidth]{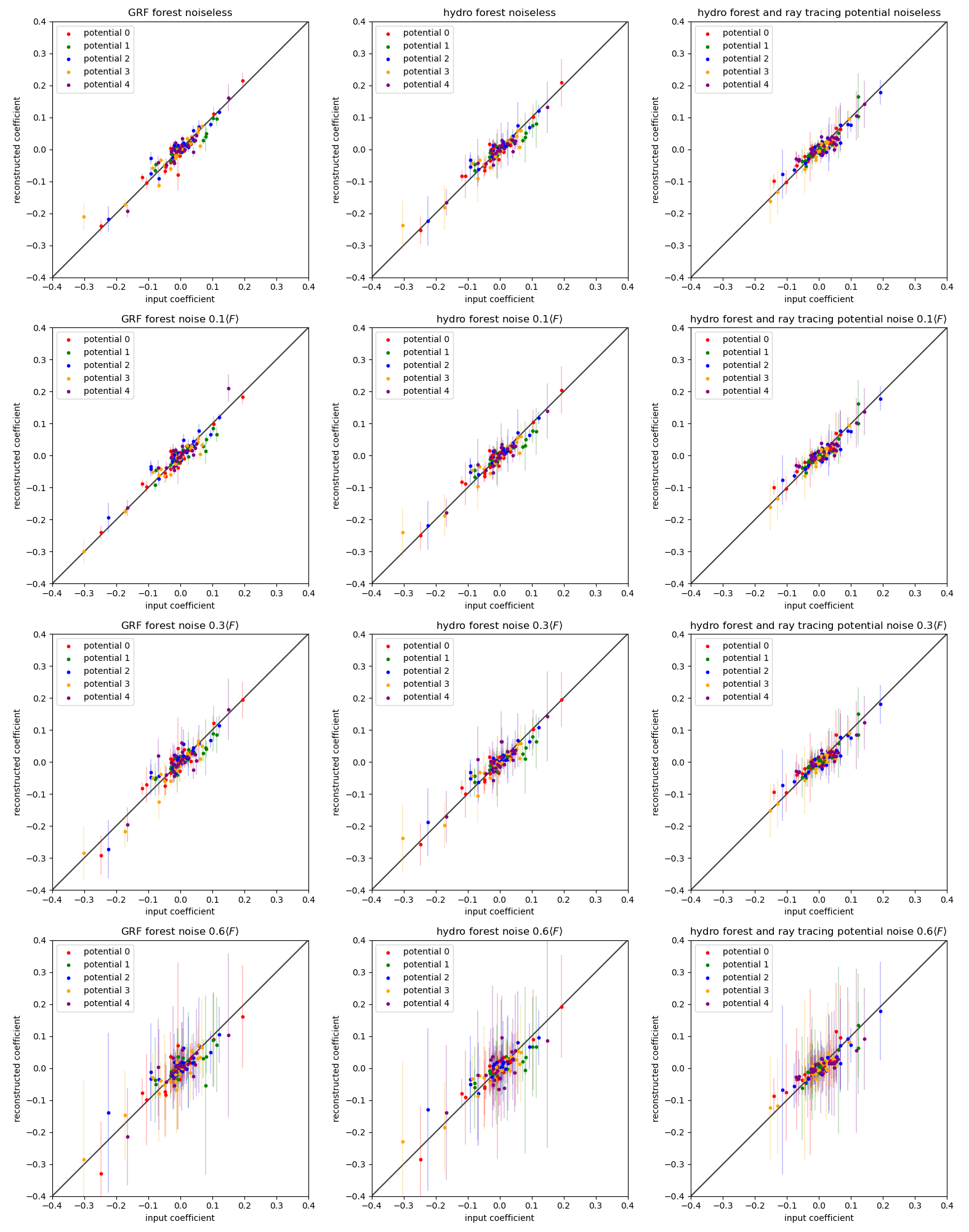}
    \caption{Scatter plots of the bias corrected reconstructed potential Legendre mode amplitudes versus the input potential Legendre mode amplitudes (in units of $10^{-6})$ for five different potentials and four different noise levels for the Gaussian forest, non-Gaussian forest, and non-Gaussian forest and input potential.  
    }
    \label{exampleConstQuasar}
\end{figure*}

\begin{figure*}
    \includegraphics[width=\textwidth]{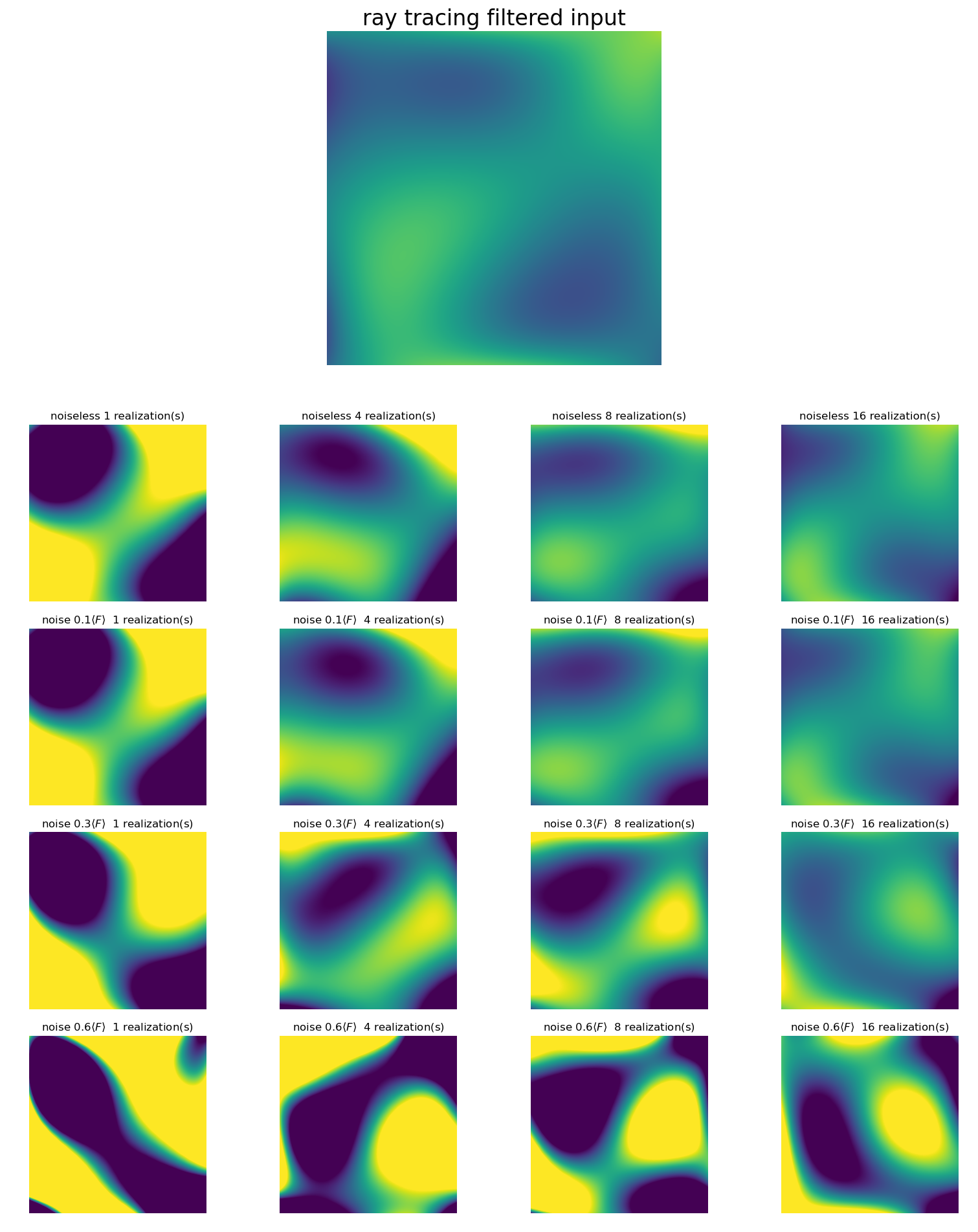}
    \caption{Lensing potential reconstructions for different amounts of data and noise. As in Figure \ref{exampleConstQuasar}, certain Legendre modes have been filtered from each potential field.
    The columns from left to right represent potentials averaged over 1, 4, 8 and 16 realizations of the \lya\ forest. Below the input row, the rows represent noise levels added to the \lya\ forest data,  at levels
    of 0, 0.1, 0.3 and 0.6 times the mean transmitted flux respectively, from top to bottom.
   As in Figure \ref{filtered potential} the field of view is a square of side length 0.655 deg, at redshift $z = 3$. }
    \label{realobs}
\end{figure*}



\begin{center}
\begin{table*}[h!]\centering
\caption{Summary of the lensing potential reconstruction statistics. The rows represent the average statistics from five different potentials at the indicated noise levels. The fit is the average slope of the input potential coefficients versus the reconstructed coefficients for five different potentials and 64 forest realizations, with error propagated from the variance of the 64 forest realizations. S/N is estimated by dividing the slope fit by the noise expected in a reconstruction from a single forest realization (standard deviation of the 64 monte carlo trials). $\chi^2$ is the reduced chi squared statistic which indicates how well the reconstruction fits the input data.}
\label{table}
\tabcolsep=0.12cm
\begin{tabular}{|c|c m{.75cm} m{.75cm}|c m{.75cm} m{.75cm}|c  m{.75cm} m{.75cm}|c  m{.75cm} m{.75cm}|} 
\multicolumn{1}{c}{}&
\multicolumn{3}{c}{GRF forest and potential}&
\multicolumn{3}{c}{hydro forest, GRF potential}&
\multicolumn{3}{c}{Gaussianized forest, GRF pot.}&
\multicolumn{3}{c}{hydro forest, ray-traced pot.}\\
\hline
noise &fit & S/N & $\chi^2$ & fit & S/N & $\chi^2$ & fit & S/N & $\chi^2$ & fit & S/N & $\chi^2$ \\
\hline
noiseless  & $ 0.73 \pm 0.02 $ & $ 2.47 $ & $ 6.58 $ & $ 0.55 \pm 0.04 $ & $ 0.91 $ & $ 2.32 $ & $ 0.6 \pm 0.03 $ & $ 1.03 $ & $ 2.37 $ & $ 0.44 \pm 0.04 $ & $ 0.69 $ & $ 2.47 $ \\
noise 0.1$\left<F\right>$   & $ 0.66 \pm 0.02 $ & $ 2.19 $ & $ 5.09 $ & $ 0.6 \pm 0.04 $ & $ 0.96 $ & $ 2.06 $ & $ 0.62 \pm 0.03 $ & $ 1.06 $ & $ 2.13 $ & $ 0.43 \pm 0.04 $ & $ 0.69 $ & $ 2.07 $ \\
noise 0.3$\left<F\right>$   & $ 0.75 \pm 0.05 $ & $ 0.9 $ & $ 1.67 $ & $ 0.7 \pm 0.06 $ & $ 0.66 $ & $ 1.17 $ & $ 0.75 \pm 0.06 $ & $ 0.74 $ & $ 1.26 $ & $ 0.5 \pm 0.06 $ & $ 0.48 $ & $ 0.9 $ \\
noise 0.6$\left<F\right>$   & $ 0.91 \pm 0.14 $ & $ 0.37 $ & $ 1.58 $ & $ 0.85 \pm 0.17 $ & $ 0.32 $ & $ 1.51 $ & $ 1.01 \pm 0.15 $ & $ 0.39 $ & $ 1.6 $ & $ 0.56 \pm 0.17 $ & $ 0.21 $ & $ 1.5 $ \\

\hline 
\multicolumn{13}{|c|}{bias corrected} \\ 
\hline

noiseless  & $ 0.86 \pm 0.03 $ & $ 2.89 $ & $ 2.02 $ & $ 0.72 \pm 0.04 $ & $ 1.16 $ & $ 0.86 $ & $ 0.71 \pm 0.04 $ & $ 1.13 $ & $ 0.93 $ & $ 0.64 \pm 0.04 $ & $ 0.97 $ & $ 0.69 $ \\
noise 0.1$\left<F\right>$   & $ 0.81 \pm 0.03 $ & $ 2.63 $ & $ 1.88 $ & $ 0.74 \pm 0.04 $ & $ 1.16 $ & $ 0.84 $ & $ 0.73 \pm 0.04 $ & $ 1.14 $ & $ 0.92 $ & $ 0.64 \pm 0.05 $ & $ 0.95 $ & $ 0.68 $ \\
noise 0.3$\left<F\right>$   & $ 0.76 \pm 0.06 $ & $ 0.91 $ & $ 0.63 $ & $ 0.73 \pm 0.07 $ & $ 0.67 $ & $ 0.33 $ & $ 0.73 \pm 0.07 $ & $ 0.66 $ & $ 0.38 $ & $ 0.64 \pm 0.08 $ & $ 0.54 $ & $ 0.27 $ \\
noise 0.6$\left<F\right>$   & $ 0.71 \pm 0.14 $ & $ 0.29 $ & $ 0.1 $ & $ 0.67 \pm 0.17 $ & $ 0.24 $ & $ 0.09 $ & $ 0.67 \pm 0.17 $ & $ 0.24 $ & $ 0.13 $ & $ 0.61 \pm 0.19 $ & $ 0.2 $ & $ 0.05 $ \\
\hline

\end{tabular}
\end{table*}
\end{center}

\subsection{Impact of non-linearity}
\label{ngforest}

In Table \ref{table}, we summarize the performance of the estimator for four different noise levels (rows) with and without applying the bias correction methods described below. At each noise level, we test the GRF simulated forest and potential, the hydrodynamic simulation forest (``hydro'') and GRF lensing potential, the hydro forest with a post hoc Gaussianization procedure (described below) and GRF potential, and the hydro forest with a lensing potential from ray tracing simulations (columns). For each case we present the best fit parameter for the slope described in Section \ref{statistics}, the signal to noise computed from the ratio of the average fit parameter to the standard deviation of the parameter fits from 64 forest realizations, and a reduced $\chi^2$ statistic weighted by the mode standard deviations to evaluate the quality of the average slope fit. 

We find that the use of hydro \lya forest flux data has an impact on the performance of the estimator (see Table \ref{table}). In the noiseless case, the hydro forest reduces the S/N of the reconstruction by a factor of $\sim 2.7$. The drop in performance becomes smaller as noise is added until the non-Gaussianity becomes negligible compared to the noise. We find that estimator appears to be biased in general as even in the Gaussian case the average slope fit is less than one.  For example, for the case with both GRF forest and
GRF lensing potential the reconstructed modes have average amplitudes that are a factor of $0.73\pm 0.02$ times the amplitudes of the modes of the input potential (top left
cell of Table \ref{table}).
The bias appears to be exacerbated by the non-linear data set as the average slope fit is even smaller in this case (reconstructed modes $0.44\pm0.04$ times
the input amplitude, top right cell of Table \ref{table}.)
We attempt a simple procedure to try to correct for this bias using our simulated data. We estimate the bias by computing the average residual for each mode across the five potentials. We find that modes seem to be biased in a consistent manner; the average residual is non-zero with statistical significance. We then correct our reconstructions by subtracting our estimated bias for each mode. This yields both a slope closer to one and more realistic error bars as evidenced by improved $\chi^2$. This bias correction method could be used even in the case of a real observation through simulated data. Explaining and managing this apparent bias is left for future work. 
The focus of this paper is evaluating the relative performance of the GRF and non-linear data sets, so the bias exhibited in the method in general is treated as a separate issue. The pixel geometry used is comparable to simulation EE in \citealt{ben1} ($512\times512$ pixels over $0.655 \times 0.655 \deg^2$ in this work compared to $500\times200$ over $0.5 \times 0.5 \deg^2$ in simulation EE). We find that our results for the Gaussian case (S/N$\sim2.5$) are consistent with those found in \citealt{ben1} (S/N of $0.67$ to $1.3$). The S/N in our case is 
larger due to sightlines that are twice as long. 

We evaluate the quality of the fit using a reduced $\chi^2$. The GRF fit is worse than the hydro case due to the smaller error bars, which are an underestimate when the bias is not corrected for. When the bias is corrected the $\chi^2$ indicates a relatively good fit. In Fig. \ref{reconstructions} we see that there is little visual discrepancy between the reconstruction from the GRF and hydro forest data and both successfully reconstruct the majority of the structure in the input potential at low to medium levels of noise.  

To mitigate the deleterious effects of non-linearity and non-Gaussianity in realistic data, we also explore the impact of a simple method of ``Gaussianizing" the non-Gaussian forest data. We rank order the \lya flux pixels and remap their values to the distribution from the Gaussian \lya simulations (see e.g., \citealt{croft98} for details). This should eliminate the non-Gaussianity at the one-point level while still maintaining the large-scale structures. We find that this procedure is successful in mitigating some of the drop in S/N (improvement of $\sim0.1$ without bias correction) and bias seen in the non-Gaussian data (see \ref{table}). The Gaussianization procedure seems to be helpful only when the reconstructions have not already been bias corrected. Further analysis is needed to determine why Gaussianization in concert with bias correction is ineffective. Both Gaussianization and bias correction could be applied to a real data set to improve estimator performance.

We also tested the estimator using both the non-linear \lya forest data and a lensing potential from ray tracing simulations. We observe that it has a similar impact to the introduction of forest non-linearity but to a lesser extent. The S/N is reduced by a factor of $\sim1.3$ for all noise levels and the bias is increased by $20-30\%$.

\subsection{Varying data quantity and forest signal to noise}
Finally, in Figure \ref{realobs} we present image reconstructions for different amounts of \lya data that include noise added  at various levels (see Section \ref{noise}). This is to allow the reader to
gain some more insight into the likely situation
for observational data from current and future
surveys.

We see that for eight realizations of the forest at low noise levels the structure of the input is well reconstructed. A substantial amount of structure is reproduced for four realizations and even one realization. One realization of the forest in this work contains $512\times512 = 262144$ \lya pixels. 
Currently available surveys such as LATIS 
 \citealt{LATIS} and CLAMATO \citealt{CLAMATO} contain $235731$ pixels in a field of similar size and $64304$ in a field of smaller size respectively, with similar source density. Therefore, the results for one forest realization should be comparable to what can be achieved from currently available data. We show that even with the impact of non-Gaussianity present in real observations, reconstruction of structure seems possible at lower noise levels. Noise is the limiting factor in currently available data sets. At realistic noise levels of 0.6 times the mean flux, some structure may still be recovered but it becomes difficult with the amount of data available. We look forward to surveys such as DESI which will contain three orders of magnitude more \lya spectra over a larger observational area. 


\section{Summary and Discussion}
\label{DiscusssionandConclusions}

\subsection{Summary}

We have further developed the field of \lya forest gravitational weak lensing by testing the performance of the \lya forest estimator of \citep{metcalf20} on more realistic data sets. We specifically evaluated the impact of the introduction of non-linearity in the both the simulated \lya forest pixel data and simulated lensing potential. As expected, deviations from Gaussianity in both the forest and lensing potential reduce the effectiveness of the estimator. The estimator was derived and proved to be optimal under the assumption of Gaussian fields (\citealt{ben1}), so we expect for more realistic fields the estimator will no longer be optimal. We find that estimator performance suffers when applied to non-linear data by a modest amount (factor of $\sim2-3$ reduction in signal to noise). However, we have presented two simple methods for mitigating the impact of non-linearity and non-Gaussianity including ``Gaussianization'' and bias correction. We find that in most cases these methods improve our results and should be applicable to real observational data. The simulated data sets used here are comparable in size to available \lya observations, although the limiting factor will likely be the noise present in observational data sets, which is at the high end of noise levels we have tested.

\subsection{Discussion}
\label{disc}
We find that the estimator appears to be biased in general, yielding systematically smaller amplitudes of reconstructed Legendre modes of the gravitational potential than those input. Some of the bias observed in the non-Gaussian case can be attributed to difficulties in accurately estimating the \lya correlation function. The estimator requires accurate a priori knowledge of the correlation function of the \lya forest in order to be unbiased. However, we observe bias even in the Gaussian case when both the estimator and forest assume the same correlation function indicating there is another source of bias. Future work will involve developing a method for more accurately estimating the correlation function of observed data and mitigating the bias present in the estimator. 

Our investigations of the signal to noise of potential detections of \lya\ forest lensing (e.g., the results
in Table \ref{table}) have involved comparisons of the true gravitational potential to the reconstructed one. This is 
complimentary to the work of \cite{metcalf20}, who  quantified  the detection 
confidence of Legendre modes with non-zero amplitude. Observationally, the true potential would not be available,
and for a comparison one would need to make an estimate, for example from the galaxy density field at the redshifts
of the lensing potential.

With present data in degree-scale surveys, such as \cite{LATIS} and \cite{CLAMATO}, we have seen that the likelihood of a detection is small, given the relatively high noise levels (S/N of order unity) in currently available \lya\ spectra. Our simulations of a non-Gaussian forest lensed by a non-Gaussian potential with a 
high, but realistic level of observational noise yield a S/N of only $0.2$ for 512 sightlines over 0.42 deg$^2$. Surveys such as DESI \cite{}, which contain hundreds of thousands to millions of \lya\ spectra
over large areas of the sky will be needed if precision cosmology with forest weak lensing is to be realised. Even with
a relatively low signal to noise detection, forest lensing will still have some advantages and differences with galaxy lensing, the most obvious being the higher source redshift (pixels at $z=2-3$), which probes the Universe at
earlier times ($z\sim0.5-1.0$).

Future work will involve testing the method with much larger, lower density simulated survey geometries similar to DESI to investigate whether a \lya weak lensing detection could be realized in this regime. We also plan to refine our methods for mitigating the bias and noise introduced by non-linearity and non-Gaussianity, uncertainty in the \lya correlation function, and intrinsic bias observed in the estimator.

\subsection*{Data availability}
The hydrodynamic simulation \lya spectra used in this work are available through request to the author.

\subsection*{Acknowledgements}

This work is supported by  NASA ATP 80NSSC18K101,  NASA ATP NNX17AK56G, NSF NSF AST-1909193, and the NSF AI Institute: Physics of the Future, NSF PHY- 2020295.


\bibliographystyle{mnras}
\bibliography{ref} 

\end{document}